# Enhanced and reduced solute transport and flow strength in salt finger convection in porous media

Xianfei Zhang(张先飞) [a,b], Ling-Ling Wang(王玲玲) [c], Shi-Di Huang(黄仕迪) [a,b,*]

[a] *Department of Mechanics and Aerospace Engineering and Center for Complex Flows and Soft Matter Research, Southern University of Science and Technology, Shenzhen 518055, China*

[b] *Guangdong Provincial Key Laboratory of Turbulence Research and Applications, Southern University of Science and Technology, Shenzhen 518055, China*

[c] *State Key Laboratory of Hydrology-Water Resources and Hydraulic Engineering, Hohai University, Nanjing 210098, China*

*Corresponding author. E-mail address: huangsd@sustech.edu.cn

## Abstract

We report a pore-scale numerical study of salt finger convection in porous media, with a focus on the influence of the porosity in the non-Darcy regime, which has received little attention in previous research. The numerical model is based on the lattice Boltzmann method with a multiple-relaxation-time scheme and employs an immersed boundary method to describe the fluid-solid interaction. The simulations are conducted in a two-dimensional, horizontally-periodic domain with an aspect ratio of 4, and the porosity $\phi$ is varied from 0.7 to 1, while the solute Rayleigh number $Ra_S$ ranges from $4 \times 10^6$ to $4 \times 10^9$. Our results show that, for all explored $Ra_S$, solute transport first enhances unexpectedly with decreasing $\phi$, and then decreases when $\phi$ is smaller than a $Ra_S$-dependent value. On the other hand, while the flow strength decreases significantly as $\phi$ decreases at low $Ra_S$, it varies weakly with decreasing $\phi$ at high $Ra_S$ and even increases counterintuitively for some porosities at moderate $Ra_S$. Detailed analysis of the salinity and velocity fields reveals that the fingered structures are blocked by the porous structure and can even be destroyed when their widths are larger than the pore scale, but become more ordered and coherent with the presence of porous media. This combination of opposing effects explains the complex porosity-dependencies of solute transport and flow strength. The influence of porous structure arrangement is also examined, with stronger effects observed for smaller $\phi$ and higher $Ra_S$. These findings have important implications for passive control of mass/solute transport in engineering applications.

## I. INTRODUCTION

Salt finger convection is a phenomenon that occurs in many natural and industrial processes when warm, salty fluids overlie cold, fresh ones. Examples include oceanic circulation, stellar interiors, magmas, and chemical engineering.[1–4] This type of convection has received extensive study due to its fundamental and practical importance, with research focusing on topics such as the onset and evolution mechanisms, characteristic scales, and transport properties.[5–11] There have also been studies exploring salt finger convection coupled with other types of flows, such as internal waves,[12] turbulent plumes and jets,[13] settling-driven flows,[14] and shear flows,[15–17] which have expanded our understanding of salt finger convection in various scenarios.

Another important area of study is salt finger convection in porous media, which is relevant to groundwater flow, metal solidification, and carbon dioxide sequestration.[18–20] Theoretical analysis suggests that there are two different flow regimes[21] when convection takes place in porous media: when the porosity is small, the flow follows Darcy's law due to viscous damping, but as the porosity becomes large, the flow enters the non-Darcy regime. Although the pioneering work on fingering convection in porous media can be dated back to at least half a century ago,[22] existing studies on this subject remain limited. As far as we know, previous studies of salt finger convection in porous media mainly focused on the Darcy regime.[23–27] Little attention has been paid to the non-Darcy case, which is the focus of the present study.

To fully understand the impact of porous media on salt finger convection, detailed information about the salinity, temperature, and velocity fields is needed. However, obtaining this information is experimentally challenging, especially given the presence of porous media and its complex interaction with the convective flow.[28] Numerical simulations, however, have become powerful tools in addressing this issue.

Previous numerical studies of salt finger convection in porous media have been carried out at a representative elementary volume scale, treating the porous media as a virtual continuous medium consisting of both solid structures and interstitial fluids.[28–29] This treatment allows for macroscopic physical quantities to be considered within an acceptable range of statistical variation, with all variables convoluted with a given spatial filter.[30–32] However, porous flows are also substantially influenced by the detailed structure of porous media.[33] Without a microscale description of the flow dynamics in pore-scale and its interaction with the porous structure, it is not possible to obtain precise information about the flow structures and associated transport properties.

In this study, we have developed a pore-scale numerical model to investigate salt finger convection in porous media. The model is based on the lattice Boltzmann method with a multiple-relaxation-time scheme. Compared to other numerical methods for studying porous flows at pore scale, e.g., direct numerical simulation and pore-network method,[34] the lattice Boltzmann method is very popular because of its advantages, such as intrinsic parallelism, ease of implementation, and ability to handle complex geometries.[35–38] Additionally, an iterative source-correction immersed boundary method was used to describe the interaction between the fingering flow and the porous structure, the accuracy of which was verified in our previous study.[39]

With this numerical model, we have conducted the first systematic study of salt finger convection in porous media in the non-Darcy regime. Our results show that the presence of porous media can either enhance or reduce the solute transport and flow strength of salt finger convection. This is in contrast to a recent study of Rayleigh–Bénard convection in porous media,[40] where heat transport efficiency was enhanced but flow strength was always reduced due to the blockage of the porous structure. As we will show in this paper, the complex behaviors of non-Darcy salt finger convection can be explained consistently by changes in flow structures.

The rest of the paper is organized as follows: in Section II, we provide a detailed description of the numerical model and methods used. In Section III, we present and discuss the effects of porous structure on the vertical transport properties and flow structures of salt finger convection, as well as examining the influence of porous structure arrangement. Finally, in Section IV, we summarize the main findings of this study.

## II. NUMERICAL MODEL AND METHODS

### A. Problem statement

In this study, we examine a two-dimensional fluid layer filled with porous media (see Fig. 1). The height and length of the computational domain are $H$ and $L = 4H$, respectively, giving it an aspect ratio of 4. This aspect ratio allows for at least four pairs of finger columns to form, even for the smallest value of the solute Rayleigh number ($Ra_S$) explored, making the statistical results robust.[41]

The top and bottom boundaries of the fluid layer are defined by impermeable and fixed plates, where the velocity of the plates is $U = 0$. The boundary conditions are no slip, with constant temperature and salinity. The top boundary is set at $T = 1$, $S = 1$, while the bottom is set at $T = 0$, $S = 0$. The temperature ($T$) and salinity ($S$) are dimensionless (see Section II.B for the nondimensionalization method). The horizontal direction has periodic boundary conditions.

The porous media is created by numerous identical solid balls with a diameter of $0.036H$, fixed inside the computational domain. The porosity ($\phi$) of the porous media is defined as $\phi = 1 - A_s/A$, where $A_s$ and $A$ are the total area of the solid skeleton and that of the entire computational domain, respectively. By varying the number of solid balls, the porosity is changed from 0.7 to 1 in this study, examining mainly non-Darcy regimes of porous flows.

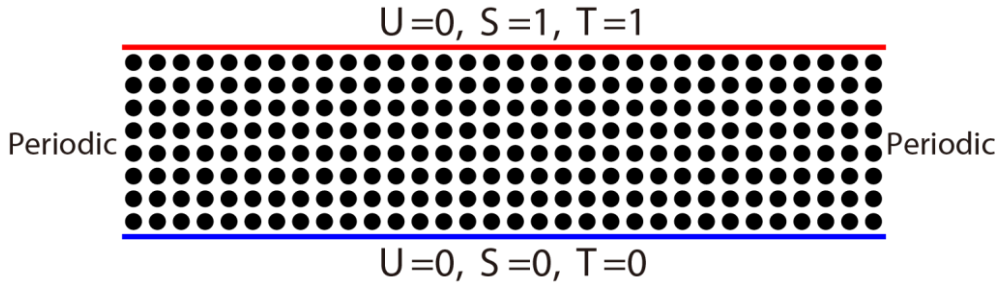

FIG. 1. Schematic of the numerical configuration. See text for a detailed explanation.

It is important to note that although salt finger convection in real-world scenarios is three-dimensional, previous numerical studies have shown that the flow evolution process and characteristic scales of flow structures are similar in two-dimensional and three-dimensional systems, with only a constant difference in their solute fluxes.[42] In a recent study by Yang *et al.*, it was also found that two-dimensional and three-dimensional simulations exhibited similar flow evolution and transport properties, leading to the conclusion that two-dimensional simulations can capture the essential dynamics of three-dimensional systems.[43] In the presence of porous media, intuition suggests that its major impact is on the horizontal scales of the fingered structures and the vertical transport properties. Thus, two-dimensional simulations can effectively reveal the dominant effects and physics. As three-dimensional simulations are still computationally intensive and prohibitive, conducting two-dimensional simulations as a first step can provide a quantitative understanding of how porous media influence salt finger convection. Future work will aim to investigate three-dimensional cases through the development of more efficient numerical methods.

**B. Governing equations and control parameters**

The fluid in the present study is incompressible and Newtonian. By applying the Boussinesq approximation, the fluid density $\rho_f$ satisfies the following formula:

$$\rho_f = \rho_0[1 - \beta_T(T - T_0) + \beta_S(S - S_0)] \tag{1}$$

where $\beta_T$ is the thermal expansion coefficient and $\beta_S$ is the salinity contraction coefficient; $\rho_0$, $T_0$, and $S_0$ are the referenced density, temperature, and salinity, respectively. Note that the density variation is only considered in the body force term in the momentum equation. In addition, because the present investigated system is convection-dominated, the Soret effect and the Dufour effect can be neglected.[44]

Based on the assumptions above, the governing equations for the mass, momentum, temperature, and salinity fields in the fluid phase and the temperature field in the solid phase are given as below:

$$\nabla \cdot \boldsymbol{u} = 0 \tag{2}$$

$$\frac{\partial \boldsymbol{u}}{\partial t} + \boldsymbol{u} \cdot \nabla \boldsymbol{u} = -\nabla p + \nu_f \nabla^2 \boldsymbol{u} + \boldsymbol{G} + Q_{\rho u} \tag{3}$$

$$\frac{\partial T}{\partial t} + \boldsymbol{u} \cdot \nabla T = \frac{\kappa_f}{(\rho_f c_f)} \nabla^2 T + Q_T \tag{4}$$

$$\frac{\partial S}{\partial t} + \boldsymbol{u} \cdot \nabla S = D_f \nabla^2 S + Q_S \tag{5}$$

$$\frac{\partial T}{\partial t} = \frac{\kappa_s}{(\rho_s c_s)} \nabla^2 T + Q_T \tag{6}$$

Here, $\boldsymbol{u} = (u, w)$ is the velocity vector with $u$ in the horizontal direction and $w$ in the vertical; $t$ represents the time; $p$ represents the pressure; and $\boldsymbol{G} = (\rho_0 - \rho)\boldsymbol{g}$ is the buoyancy force with $\boldsymbol{g}$ being the gravitational acceleration vector. The physical properties include the molecular viscosity $\nu$, the thermal conductivity $\kappa$, the specific heat capacity $c$, and the salinity diffusivity $D_f$. Note that the subscripts $*_f$ and $*_s$ denote the properties for the fluid and solid phases, respectively.

The interaction between the solid porous structure and the flow is described using the immersed boundary method, which requires the inclusion of additional terms in the governing equations. These terms are referred to as the body force $\boldsymbol{Q}_{\rho u}$, heat source $Q_T$, and salinity source $Q_S$. The boundary conditions at the surface of the solid porous structure are defined by the following mathematical expressions: no slip [Eq. (7a)], no salinity flux [Eq. (7b)], and fluid-solid conjugate heat transfer [Eqs. (7c) and (7d)], respectively.

$$\boldsymbol{u} = 0 \tag{7a}$$

$$\frac{\partial S}{\partial n} = 0 \tag{7b}$$

$$T_f = T_s \tag{7c}$$

$$\kappa_f \frac{\partial T_f}{\partial n} = \kappa_s \frac{\partial T_s}{\partial n} \tag{7d}$$

To nondimensionalize the governing equations, the system height $H$ and the thermal diffusion time $\frac{H^2}{\kappa_f/(\rho c)_f}$ are used as the characteristic length and time, respectively. The characteristic temperature and salinity are defined as the temperature and salinity differences between the top and bottom boundaries, i.e., $\Delta_T = T_{\text{top}} - T_{\text{bottom}}$ and $\Delta_S = S_{\text{top}} - S_{\text{bottom}}$. This leads to the following four control parameters: the Prandtl number $Pr = \frac{\nu_f}{\kappa_f/(\rho c)_f}$, the Lewis number $Le = \frac{\kappa_f/(\rho c)_f}{D_f}$, the density ratio $N = \frac{\beta_S \Delta_S}{\beta_T \Delta_T}$, and the solute Rayleigh number $Ra_S = \frac{g \beta_S \Delta_S H^3}{(\nu D)_f}$. In this study, $Pr = 7$, $Le = 100$, and $N = 2$ are fixed, while the $Ra_S$ number is varied from $4 \times 10^6$ to $4 \times 10^9$. Notably, the range of $Ra_S$ explored is coupled with the porosity range ($0.7 \leq \phi \leq 0.1$) as the focus is on the non-Darcy regime. As this paper will show, a steep drop in flow strength occurs when $Ra_S < 4 \times 10^6$ and $\phi < 0.7$, which signals a transition to the Darcy regime and is outside the scope of the present study.

**C. Numerical methods and grid independence test**

In this study, an Immersed Boundary lattice Boltzmann method is utilized to solve the fluid layer filled with porous media. This method has been demonstrated to be efficient, accurate, and stable in simulating porous flows.[44] To ensure stability in solving the solute transport equation, a multiple-relaxation-time scheme is adopted.[45] An iterative source-correction immersed boundary method is utilized to describe the fluid-structure interaction. For more information on the numerical methods, please refer to a previous publication,[39] where the accuracy of the model in simulating salt finger convection in porous media was demonstrated.

To ensure the accuracy of the numerical results, a grid independence test was conducted at $Ra_S = 4 \times 10^9$ and $\phi = 0.87$ using three different grid resolutions: $802 \times 201$ (low-resolution), $1602 \times 401$ (moderate resolution), and $2402 \times 601$ (high-resolution). The low-resolution case did not produce a stable solution. However, the moderate resolution and high-resolution cases showed little difference in their time series of solute Nusselt number ($Nu_S(t)$) and flow structures, as shown in Fig. 2.

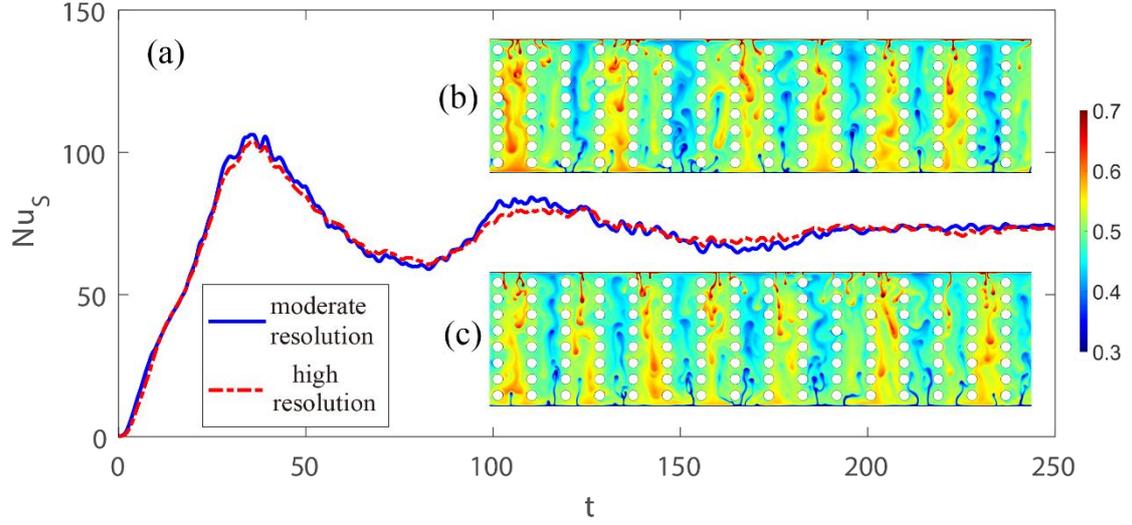

FIG. 2. (a) Time series of solute Nusselt number $Nu_S(t)$ for different numbers of grids at $Ra_S = 4 \times 10^9$ and $\phi = 0.87$. The insets show typical snapshots of the salinity field at the quasi-steady stage: (b) moderate resolution; (c) high-resolution.

We evaluated the two most important quantities in this study: the time-averaged solute Nusselt number $Nu_S$ and flow strength $Re_z$ after reaching the quasi-steady stage. Here, $Nu_S$ is defined by $Nu_S = -\langle wS \rangle_V / (D_f \Delta_S H^{-1})$, where $\langle \cdot \rangle_V$ represents spatial-averaging over the entire computational domain, and $Re_z = w_{rms} H / \nu_f$, where $w_{rms}$ is the root-mean-square value of the vertical velocity. The results showed that $Nu_S$ and $Re_z$ are 73.8 and 12.7 for the moderate-resolution case, and 73.3 and 12.5 for the high-resolution case, with relative errors of 0.7% and 1.6%, respectively. These results indicate that the moderate resolution is sufficient for solving the present problem, and the moderate-resolution solution with a total of $1602 \times 401$ grids was adopted for the study.

## III. RESULTS AND DISCUSSION

### A. Flow evolution

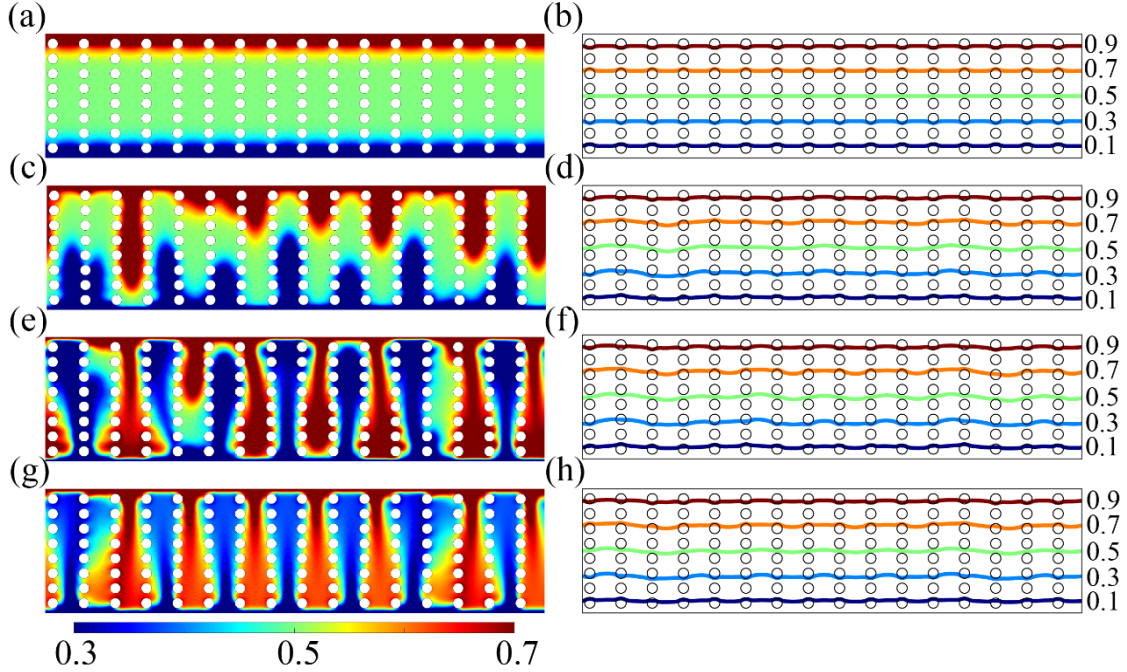

FIG. 3. Typical snapshots of the distribution of salinity (left panel) and isotherms (right panel) of salt finger convection in porous media for $\phi = 0.87$ and $Ra_S = 4 \times 10^6$ at four moments: (a, b) $t = 200$, (c, d) $t = 500$, (e, f) $t = 600$, and (g, h) $t = 2000$. The color bar in the left panel represents the salinity magnitude and the isotherms are indicated by the values in the right panel.

Figure 3 illustrates the salinity fields and isotherms of salt finger convection in porous media for $\phi = 0.87$ and $Ra_S = 4 \times 10^6$. Four snapshots taken at different stages of the evolution process are shown, including the diffusion-dominant stage, the finger development stage, the mutual mixing stage, and the quasi-steady stage. During the diffusion-dominant stage [Fig. 3(a)], the formation of fingered columns has not yet begun, and transport properties are primarily driven by molecular diffusion. At the finger development stage [Fig. 3(c)], fluid near the upper and lower boundaries becomes unstable and begins to form fingered structures, which are restricted by the porous media's solid structure as they move along the pores. As time passes, the up- and down-going columns reach each other and the system enters the mutual mixing stage [Fig. 3(e)], where the columns interact, causing twisting and swaying motions. Finally, fully developed fingered columns are formed at the quasi-steady stage [Fig. 3(g)]. Notably, throughout the entire evolution process, the temperature field remains nearly linearly distributed in the vertical direction, as seen in the right panel of Fig. 3.

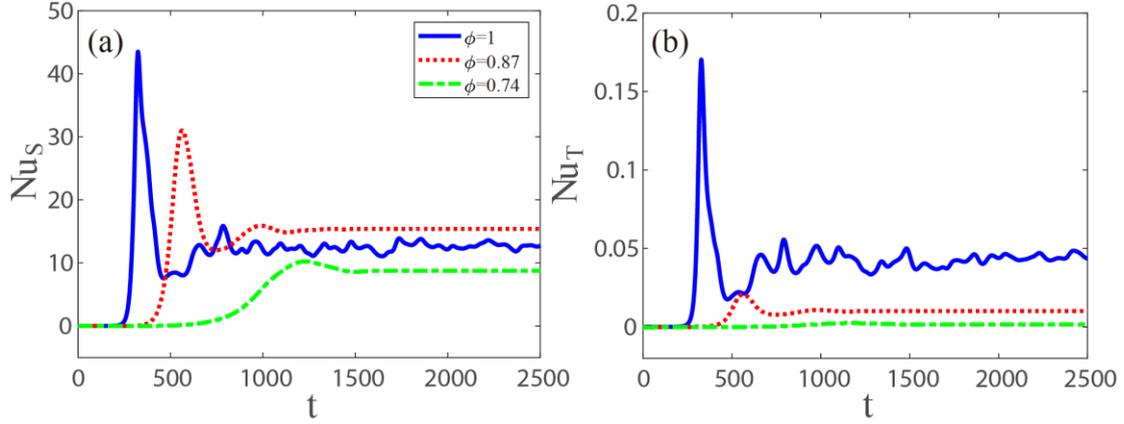

FIG. 4. Time series of (a) solute Nusselt number $Nu_S(t)$ and (b) thermal Nusselt number $Nu_T(t)$ for different porosities at $Ra_S = 4 \times 10^6$.

The evolution of salt finger convection can also be observed in Fig. 4, which shows the time series of the nondimensional salinity flux $Nu_S(t)$ and heat flux $Nu_T(t)$ for different values of $\phi$ at $Ra_S = 4 \times 10^6$. Here, $Nu_S(t)$ and $Nu_T(t)$ are defined by $Nu_S = -\langle wS \rangle_V / (D_f \Delta_S H^{-1})$ and $Nu_T = -\langle wT \rangle_V / [\kappa_f/(\rho_f c_f) \Delta_T H^{-1}]$, respectively, where $\langle \cdot \rangle_V$ represents the spatial average over the entire computational domain. As seen in Fig. 4(a), the salinity flux curves for different porosities follow a similar evolution trend. The salinity flux stays at a level of zero for a finite time, then starts to increase and reaches a maximum value, before decreasing and eventually reaching a quasi-steady value. Similar patterns can be seen in the nondimensional heat flux $Nu_T(t)$ displayed in Fig. 4(b). These variations in the transport properties align with the four stages of salt finger evolution described in Fig. 3.

It can also be observed in Fig. 4 that the finger development is delayed with the presence of porous media, and the delay becomes more pronounced with decreasing porosity. Furthermore, the fluctuations in $Nu_S(t)$ and $Nu_T(t)$ disappear when the porosity is less than 1. This is because the finger structures can move freely with twisted and swayed motions in the absence of porous media. Meanwhile, the upwelling and down-going columns can interact with each other directly. Both dynamical processes contribute to the small fluctuations in salinity and heat fluxes, as seen in Fig. 4. In the presence of porous media, the flow is not free owing to the restriction of the porous media, leading to the formation of more ordered structures between the pores. As a result, the fluctuations are weakened or suppressed for cases with lower $Ra_S$ numbers (weaker driving force) and smaller porosity (stronger blocking effect). For cases with larger $Ra_S$ number, fluctuations in $Nu_S(t)$ and $Nu_T(t)$ can still be observed when the porosity is less than 1 (see Fig. 2).

The most interesting finding in Fig. 4 is that while heat transport decreases monotonically with decreasing $\phi$, salinity transport shows an unexpected behavior. As seen in Fig. 4(a), when the flow reaches a quasi-steady state, the salinity flux at $\phi =$ 0.87 is larger than that at $\phi = 1$. Given that heat transport in this system is negligible compared to salinity transport (at least two orders of magnitude smaller), the focus will be on the salinity transport in the following discussions to understand this unexpected behavior.

**B. Solute transport and flow strength**

Figure 5(a) shows the time-averaged value of $Nu_S$ as a function of $Ra_S$ for different porosities $\phi$, which are obtained after reaching the quasi-steady stage. For salt finger convection without porous media (i.e., $\phi = 1$), $Nu_S$ increases with $Ra_S$ and follows a power law, $Nu_S \sim Ra_S^{0.22}$, which agrees well with previous findings in a three-dimensional system.[46] However, when $\phi < 1$, no simple power law can be detected. Most of the results show an enhancement in $Nu_S$ compared to the case with $\phi = 1$, with only a few exhibiting a reduction. To better understand the effect of porosity on $Nu_S$, the data are normalized by using the values at $\phi = 1$ as a reference. The results, shown as a function of $\phi$ in Fig. 5(b), show a similar trend for all explored $Ra_S$. The data first increase with decreasing $\phi$, with an enhancement of up to 40%. Then, the data decrease as $\phi$ is further decreased, but the magnitude of $Nu_S(\phi)$ for most cases is still larger than $Nu_S(\phi = 1)$.

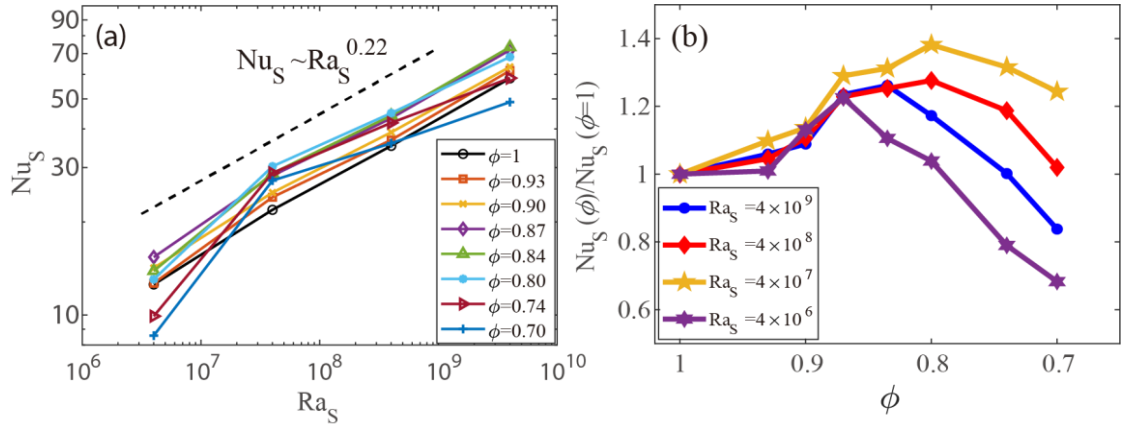

FIG. 5. (a) $Nu_S$ as a function of $Ra_S$ for different $\phi$. (b) $Nu_S(\phi)/Nu_S(\phi = 1)$ as a function of $\phi$ for different $Ra_S$. The dashed line in (a) manifests a power law of $Nu_S \sim Ra_S^{0.22}$ and it is used to guide the eye.

This nonmonotonic behavior of $Nu_S(\phi)$ is similar to phenomena observed in Rayleigh–Bénard convection in porous media and other convection systems subject to stabilizing mechanisms (e.g., spatial confinement and rotation),[40,47–55] (see details in a recent review [56]). In these systems, referred to as stabilizing-destabilizing (S-D) flows in the literature,[50] significant heat transport enhancement can be observed even though the flow strength is reduced due to stabilizing mechanisms, with the enhancement generally increasing monotonically with the Rayleigh number. However, in this study, the enhanced salinity transport induced by the porous structure (the stabilizing mechanism of the system being studied here) and the corresponding optimal porosity for salinity transport enhancement depend on $Ra_S$ nonmonotonically. Specifically, the maximum enhancement in salinity transport occurs at $Ra_S = 4 \times 10^7$, which is neither the highest nor the lowest $Ra_S$ in this study. Furthermore, the value of $\phi$ for maximum $Nu_S$ enhancement decreases from 0.87 to 0.80 as $Ra_S$ increases from $4 \times 10^6$ to $4 \times 10^7$ and $4 \times 10^8$, then increases back to 0.84 at $Ra_S = 4 \times 10^9$. These results suggest that salt finger convection in porous media is not a simple S-D flow.

The flow strength, characterized by the Reynolds number $Re_z = w_{rms} H/\nu$, shows an even more complex dependence on the porosity. In this study, the root-mean-square value of the vertical velocity at the quasi-steady stage ($w_{rms}$) is used as the characteristic velocity, as the focus is on the vertical transport properties and there is no well-defined mean flow. Note that similar results are obtained when the velocity magnitude $|U| = \sqrt{u^2 + w^2}$ is used to calculate the Reynolds number. Figure 6(a) shows the dependence of the $Re_z$ on $Ra_S$ for different porosities. For the case without porous media, the $Re_z(Ra_S)$ follows a power law of $Re_z \sim Ra_S^{0.38}$, which is in good agreement with previous findings from a study of three-dimensional salt finger convection.[46] It is observed that the $Re_z$ at $Ra_S = 4 \times 10^6$ is greatly reduced as the porosity decreases when $\phi < 1$, as seen more clearly in Fig. 6(b), where the data at $\phi = 1$ are taken as a reference. However, in contrast to the significant drop at $Ra_S = 4 \times 10^6$, the $Re_z$ at $Ra_S = 4 \times 10^9$ only varies weakly with $\phi$, and even increases significantly for some porosities at intermediate $Ra_S$. These trends in $Re_z(\phi)$ are different from previous findings in S-D flows, where the flow strength always decreases with increasing stabilization levels.[56] To understand these complex trends in $Re_z(\phi)$ and $Nu_S(\phi)$, we examine the salinity and velocity fields in detail.

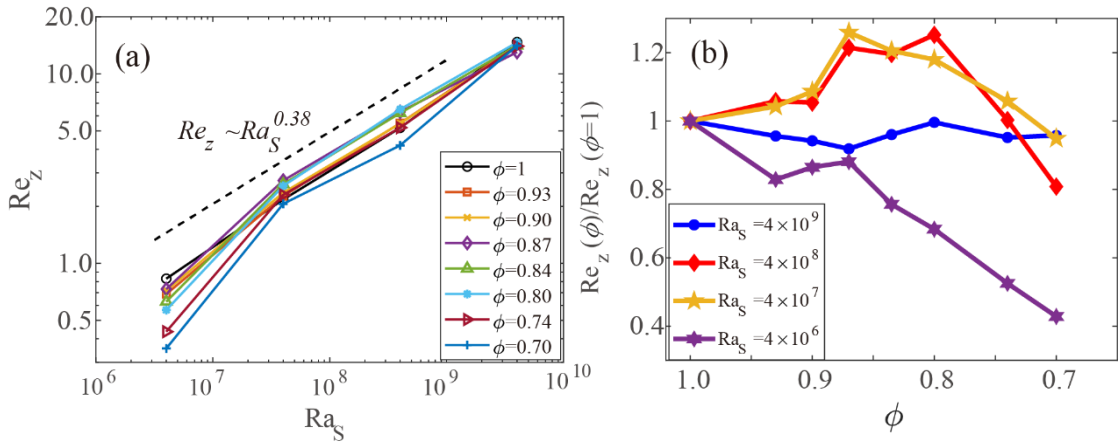

FIG. 6. (a) $Re_z$ as a function of $Ra_S$ for different $\phi$. (b) $Re_z(\phi)/Re_z(\phi = 1)$ as a function of $\phi$ for different $Ra_S$. The dashed line in (a) manifests a power law of $Re_z \sim Ra_S^{0.38}$ and it is used to guide the eye.

## C. Flow structures and flow coherence

The salinity difference drives salt finger convection; therefore, we first examine how the salinity field is impacted by the porous media. Figure 7 displays typical snapshots of the salinity field at the quasi-steady stage for three porosities and two $Ra_S$ numbers. For $Ra_S = 4 \times 10^9$, it is evident that the domain without porous media is filled with irregular and distorted fingered structures. However, the presence of porous media restricts the fluid's movement, resulting in the formation of more ordered and vertically-extended structures between the pores. This indicates that the salt fingers retain more salinity and become more coherent as they travel up and down in the presence of porous media. The porous media hinders the fluid's movement but weakens the interaction between ascending and descending salt fingers and their entrainment with the background fluid, which increases the flow coherence. As a result, some of the fingered structures become more energetic and travel straight through the domain, but the porous structure reduces their number, particularly for the case with $\phi = 0.74$.

For $Ra_S = 4 \times 10^6$, similar changes can be found; however, there are two differences. First, the fingered structures at $\phi = 1$ are much less but more ordered compared to the those at $Ra_S = 4 \times 10^9$. In addition, their number first increases as $\phi$ is decreased from 1 to 0.87, and then decreases with $\phi$ further decreasing, exhibiting a nonmonotonical behavior. Second, in contrast to the cases at $Ra_S = 4 \times 10^9$, where the salt fingers are always smaller than the pore scales of the porous media, the fingered structures at $Ra_S = 4 \times 10^6$ are comparable to and even larger than the pore sizes. This implies that the porous media have a strong blockage effect on the flow at $Ra_S = 4 \times 10^6$.

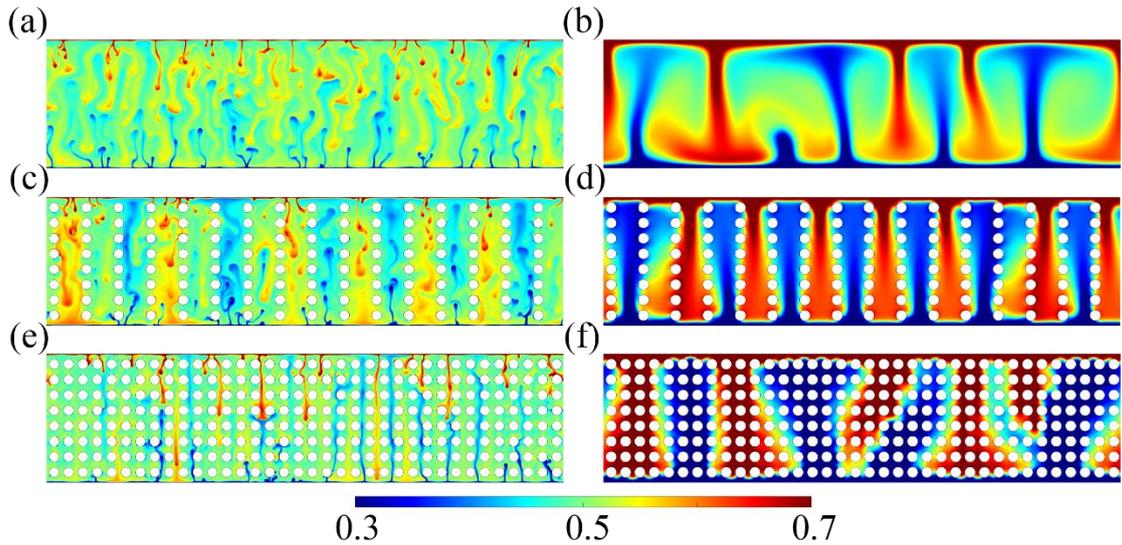

FIG. 7. Typical snapshots of the salinity field at the quasi-steady stage for different $\phi$: (a, b) $\phi = 1$, (c, d) $\phi = 0.87$, and (e, f) $\phi = 0.74$. Left panel: $Ra_S = 4 \times 10^9$; right panel: $Ra_S = 4 \times 10^6$. The color represents the salinity magnitude.

To demonstrate the impact of the porous media on the fingered structures, Fig. 8 shows the superimposed streamlines on the corresponding salinity fields depicted in Fig. 7. For clarity, only the middle section of the domain is displayed. It can be seen that as the porosity is reduced, the streamlines at $Ra_S = 4 \times 10^9$ become more regular and vertically-aligned, which is consistent with the changes in the salinity field. These vertically aligned streamlines eventually form straight upwelling and downwelling channels that connect the top and bottom boundaries directly. On the other hand, the porous media block some of the flow, leading to the formation of small vortices in between the pores for smaller porosities. This blocking effect is much stronger at $Ra_S = 4 \times 10^6$. When the porous structure is present, the domain-sized vortices at $\phi = 1$ break down into many pore-scale vortices. Additionally, while the upwelling and downwelling channels are prominent at $\phi = 0.87$, they are disrupted by the porous structure when the porosity is further reduced to 0.74. As a result, some of the streamlines in the channels become inclined, indicating a less efficient vertical motion. These results highlight that the influence of the porous media on salt finger convection is twofold: it can block or even destroy the flow structures, but it can also enhance the flow coherence and encourage the flow motion. The interplay of these opposing effects determines how the flow strength varies with porosity.

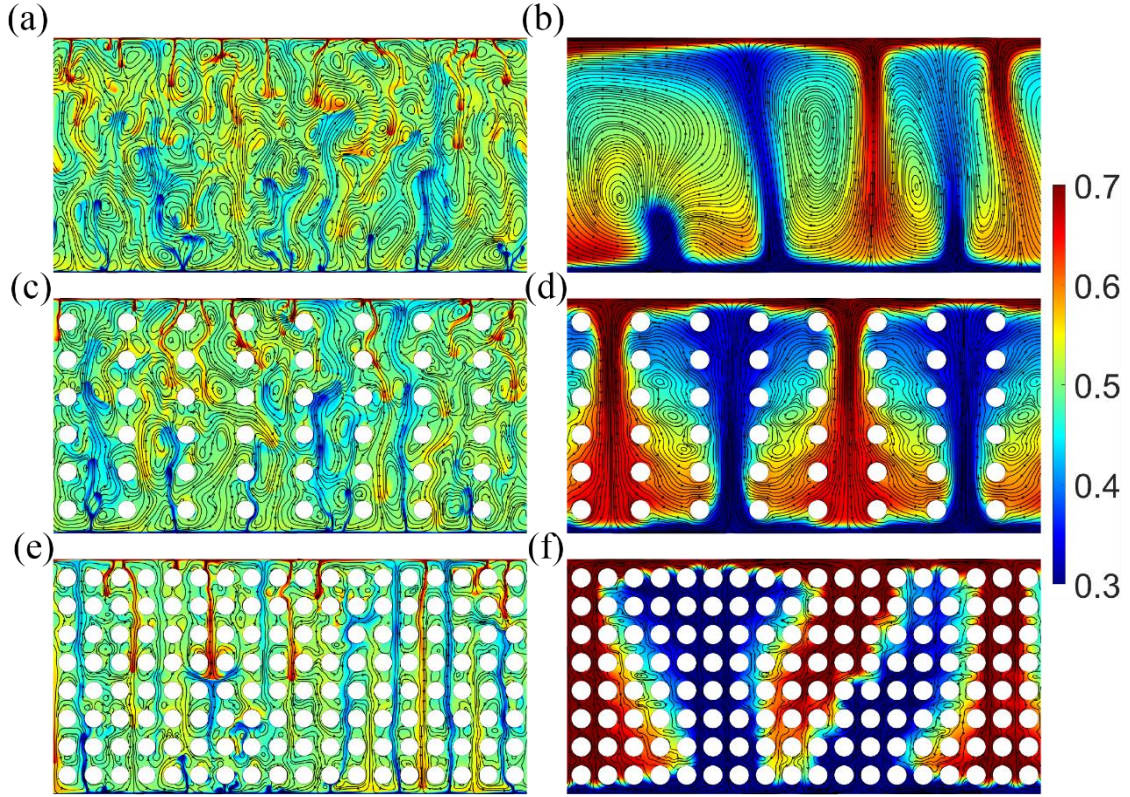

FIG. 8. Typical snapshots of the streamlines superposed on the salinity fields at the quasi-steady stage for different $\phi$: (a, b) $\phi = 1$, (c, d) $\phi = 0.87$, and (e, f) $\phi = 0.74$. Left panel: $Ra_S = 4 \times 10^9$; right panel: $Ra_S = 4 \times 10^6$. The background color represents the salinity magnitude.

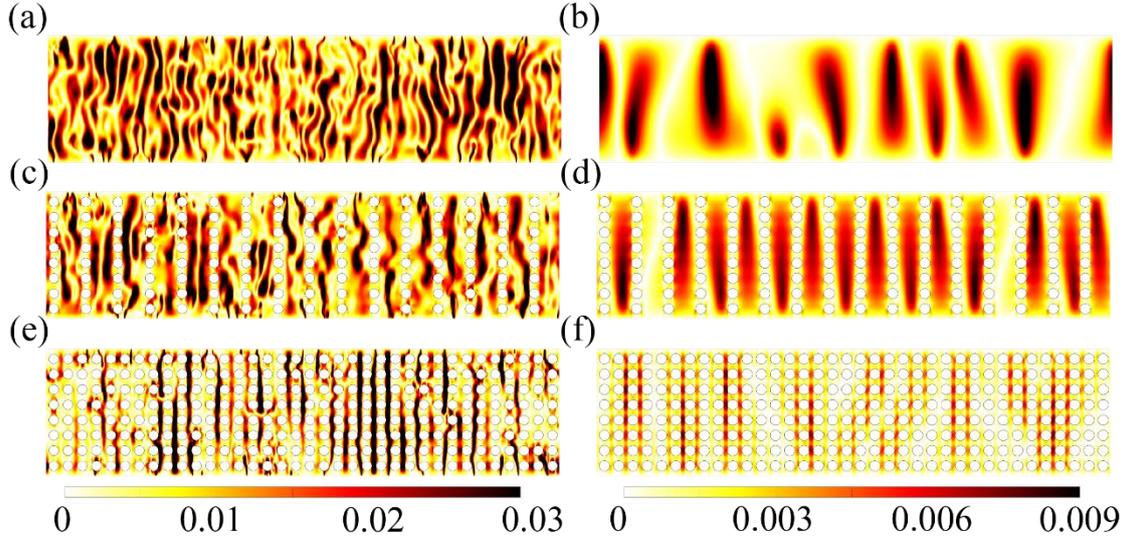

FIG. 9. Typical snapshots of the vertical velocity field at the quasi-steady stage for different $\phi$: (a, b) $\phi = 1$, (c, d) $\phi = 0.87$, and (e, f) $\phi = 0.74$. Left panel: $Ra_S = 4 \times 10^9$; right panel: $Ra_S = 4 \times 10^6$. The velocity magnitude is indicated by the color.

To better understand the factors that influence the variation trends of $Re_z(\phi)$ for different $Ra_S$ [as seen in Fig. 6(b)], we can look at the combined vertical velocity fields shown in Fig. 9 (which correspond to the salinity fields in Fig. 7). For $Ra_S = 4 \times 10^9$, the flow is full of finger structures; therefore, their number decreases immediately when there is porous media. However, this blockage effect is offset by an increase in flow coherence. As a result, the magnitude of the vertical velocity changes only slightly with $\phi$ at $Ra_S = 4 \times 10^9$. For $Ra_S = 4 \times 10^6$, the finger structures are roughly the same size as, or even larger than, the pore scales; therefore, the blockage effect becomes dominant. As seen in Fig. 9(f), the flow at $\phi = 0.74$ is so strongly blocked by the porous media that only a small portion of fluid is moving vertically. As a result, the flow strength decreases significantly as $\phi$ decreases at this $Ra_S$. It is worth noting that the number of finger columns at $Ra_S = 4 \times 10^6$ first increases with $\phi$ and then decreases, which could be responsible for the local maximum of $Re_z$ observed at $\phi = 0.87$ [as seen in Fig. 6(b)].

The two conflicting effects of the porous media can also explain the counterintuitive increase in $Re_z(\phi)$ for moderate $Ra_S$. Taking $Ra_S = 4 \times 10^7$ as an example, it is seen in Fig. 10 that the finger structures become more ordered and coherent in the presence of porous media. When $\phi$ changes from 1 to 0.87, the structures become coherent enough to connect the top and bottom boundaries directly, but their number remains unchanged. Thus, the flow strength becomes stronger due to the enhanced coherence. When $\phi$ is further reduced to 0.74, the blockage effect becomes more pronounced, causing the flow to slow down. Based on these two influences of the porous media on salt finger convection, the complex behavior of $Re_z(\phi)$ is consistently explained.

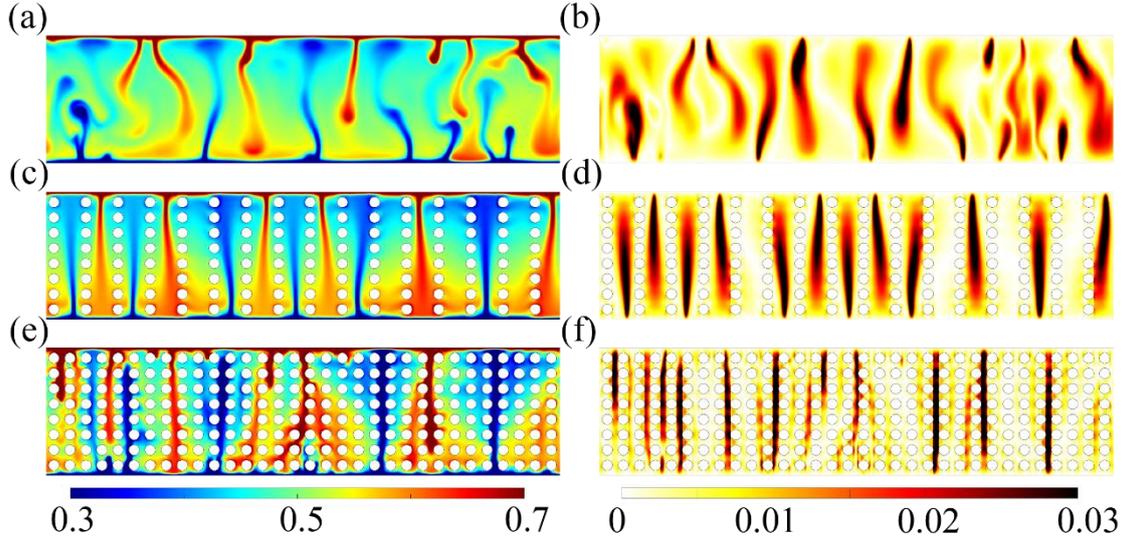

FIG. 10. Typical snapshots of the salinity field (left panel) and the vertical velocity field (right panel) at the quasi-steady stage for $Ra_S = 4 \times 10^7$ with different $\phi$: (a, b) $\phi = 1$, (c, d) $\phi = 0.87$, and (e, f) $\phi = 0.74$. The salinity and velocity magnitudes are indicated by the corresponding color bars, respectively.

In the following, we analyze the nonmonotonic behavior of $Nu_S(\phi)$ as seen in Fig. 5. To understand this behavior, we examine the correlation between the local salinity contrast $S'$ and vertical velocity $w$. Figure 11 shows the joint probability density functions (PDFs) of $S'$ and $w$ for different $\phi$ and $Ra_S$, which have been averaged over the entire domain. The distributions of $P(w, S')$ are primarily found in the second and fourth quadrants, which reflect the basic nature of salt finger convection: saltier columns move downward while fresher columns move upward. At lower $Ra_S$, the flow is more ordered and coherent, as seen in Figs. 7–9, and the distribution of $P(w, S')$ is more compact. With the presence of porous media, the proportion of large vertical velocities decrease, while the proportion of large salinity contrasts increases, especially at $Ra_S = 4 \times 10^6$. This suggests that, although porous media hinder the vertical flow motion, they weaken the interaction and entrainment between ascending and descending fingered structures with the background fluid, leading to weaker mixing in the salinity field. These results demonstrate that the correlation between $S'$ and $w$ is an effective indicator of changes in flow coherence.

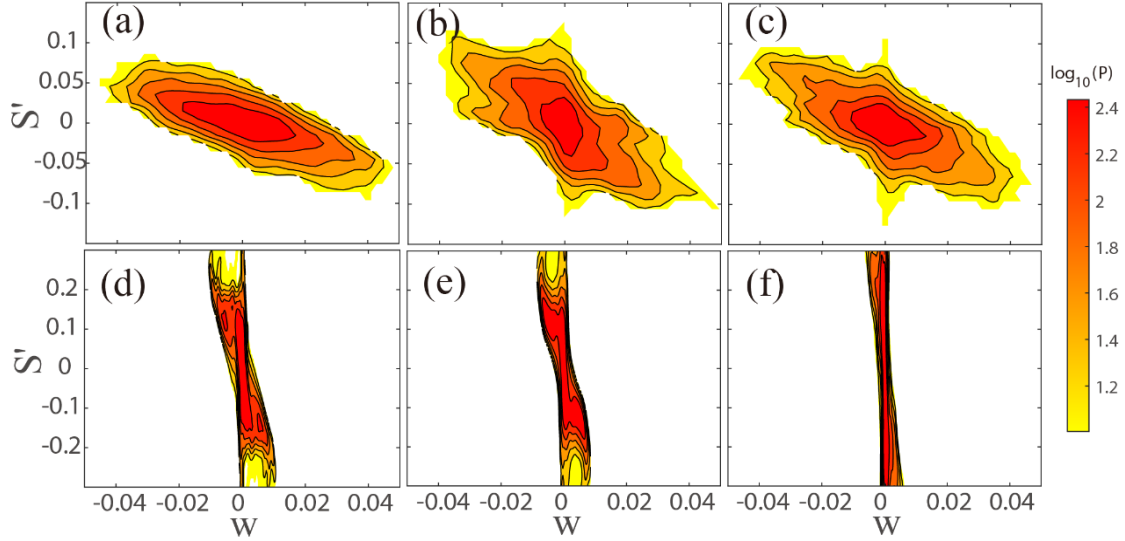

FIG. 11. Joint PDFs of the vertical velocity $w$ and the local salinity contrast $S' = S - \bar{S}$ for different $\phi$: (a, b) $\phi = 1$, (c, d) $\phi = 0.87$, and (e, f) $\phi = 0.74$. Top panel: $Ra_S = 4 \times 10^9$; bottom panel: $Ra_S = 4 \times 10^6$.

To quantify the flow coherence, we calculate the correlation between $S'$ and $w$ by $C \equiv \langle \varphi(S') \cdot \varphi(w) \rangle$, where $\varphi(X) \equiv [X - \langle X \rangle]/\sigma_X$ with $X$ being the corresponding variables. A high value of $C$ indicates that the vertical velocity is well correlated with the salinity contrast, meaning increased flow coherence generally helps with salinity transport while decreased flow coherence reduces it. To directly reflect the influence of porous media, the data at $\phi = 1$ are taken as the reference. The normalized correlation value $|C(\phi)|/|C(\phi = 1)|$ is plotted in Fig. 12(a) and shows that for all $Ra_S$, the flow coherence increases and then decreases as $\phi$ is decreased, with the most significant enhancement occurring at $Ra_S = 4 \times 10^6$. This provides direct, quantitative evidence that salt finger convection becomes more coherent when there is porous media.

However, the correlation value does not reflect the change in flow intensity, which also plays a significant role in solute transport. (Note that a highly coherent flow does not necessarily mean a stronger flow.) To account for both flow intensity and coherence, we further define $F = Re_z \cdot |C|$ as the effective transport coefficient. The magnitude of $F$ quantifies the salinity transport capacity, and using the data at $\phi = 1$ as a reference, the normalized value of $F(\phi)/F(\phi = 1)$ indicates whether solute transport is increased or decreased compared to the case without porous media. Figure 12(b) shows that the normalized effective transport coefficient $F(\phi)/F(\phi = 1)$ largely captures the overall variation trend of $Nu_S(\phi)/Nu_S(\phi = 1)$ in Fig. 5(b). The similarity between the behaviors of $F(\phi)/F(\phi = 1)$ and $Nu_S(\phi)/Nu_S(\phi = 1)$ allows us to arrive at the following understanding of salinity transport in salt finger convection in porous media.

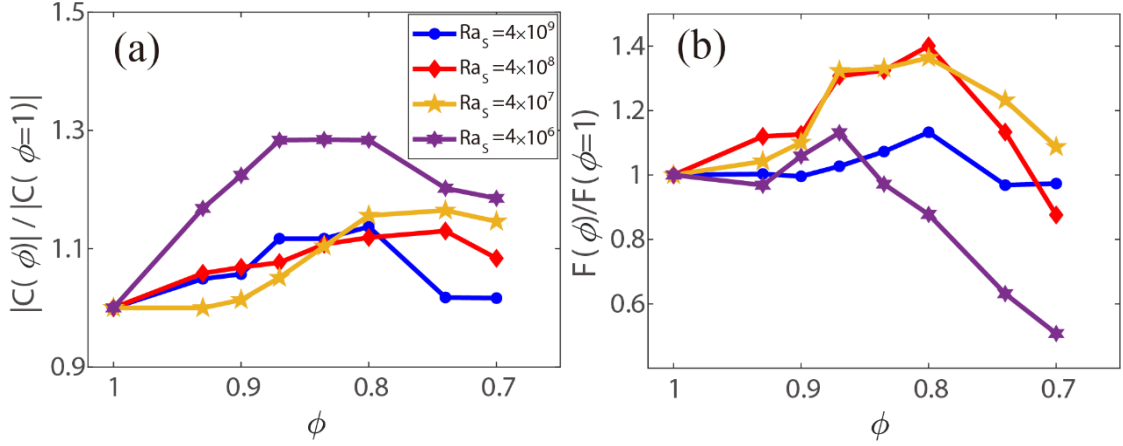

FIG. 12. (a) Normalized correlation value $\frac{|C(\phi)|}{|C(\phi=1)|}$ and (b) effective transport coefficient $\frac{F(\phi)}{F(\phi=1)}$ as a function of $\phi$ for different $Ra_S$.

For low solute Rayleigh numbers $Ra_S$, the porous media disrupts the dominant flow structure of salt finger convection, resulting in slower motion. However, the flow coherence is greatly improved due to reduced flow fluctuations and mixing. As a result, salinity transport $Nu_S$ is enhanced for some porosities when the blockage effect on the flow is not dominant, a finding similar to that in Rayleigh–Bénard convection in porous media.[40]

At a moderate level of $Ra_S$, the fingered structures become thin and are not significantly blocked by the porous structure in the porosity range explored. As a result, the increase in flow coherence becomes the dominant influence, leading to enhancements in both flow strength and salinity transport. The increase in flow intensity further enhances salinity transport. This is why the maximum enhancement in $Nu_S$ occurs at moderate $Ra_S$ rather than the highest or lowest $Ra_S$ as seen in previous studies on S-D flows.[56]

For higher $Ra_S$ values, as previously discussed, the flow strength varies weakly with porosity ($\phi$). Thus, the enhancement in $Nu_S$ is mainly due to the increase in flow coherence. In short, the difference between salt finger convection in porous media and S-D flows lies in the characteristics of the dominant flow structures in the system, which determine how the flows respond to different stabilization mechanisms.

**D. Effect of the porous structure arrangement**

In previous sections, the porous media was arranged regularly in the domain. In this section, we delve deeper into the impact of the porous structure arrangement. We consider three common arrangements: lattice, staggered, and random (as shown in Fig. 13). The random arrangement has a minimum pore scale $d_{\min}$ that satisfies $d_{\min} \geq 0.02H$, which is resolved by at least eight grids to ensure accurate flow dynamics. We conduct simulations with three representative porosities across the $Ra_S$ range of $4 \times 10^6 \leq Ra_S \leq 4 \times 10^9$.

Figure 13 demonstrates how the porous structure arrangement impacts the flow structures at $Ra_S = 4 \times 10^7$. When the porosity is relatively high ($\phi = 0.93$), the flow patterns are similar in all arrangements, characterized by fingered structures that rise and fall with limited interference from the porous media. However, at $\phi = 0.87$, the effect of the porous structure arrangement becomes significant. The regular vertical convection channels become zigzag and even form multi-forked, tree-like structures when the porous structure changes from lattice to staggered or random arrangements. For $\phi = 0.74$, the changes in the flow pattern caused by different porous structure arrangements are similar to the case with $\phi = 0.87$. However, because the pore scales are smaller than the fingered structure widths at $\phi = 0.74$, the vertical channels are already obstructed by the porous media in the case with lattice arrangement. A similar phenomenon can be observed at $Ra_S = 4 \times 10^9$, but the difference in flow structures is much greater as the porosity decreases (refer to Fig. 14).

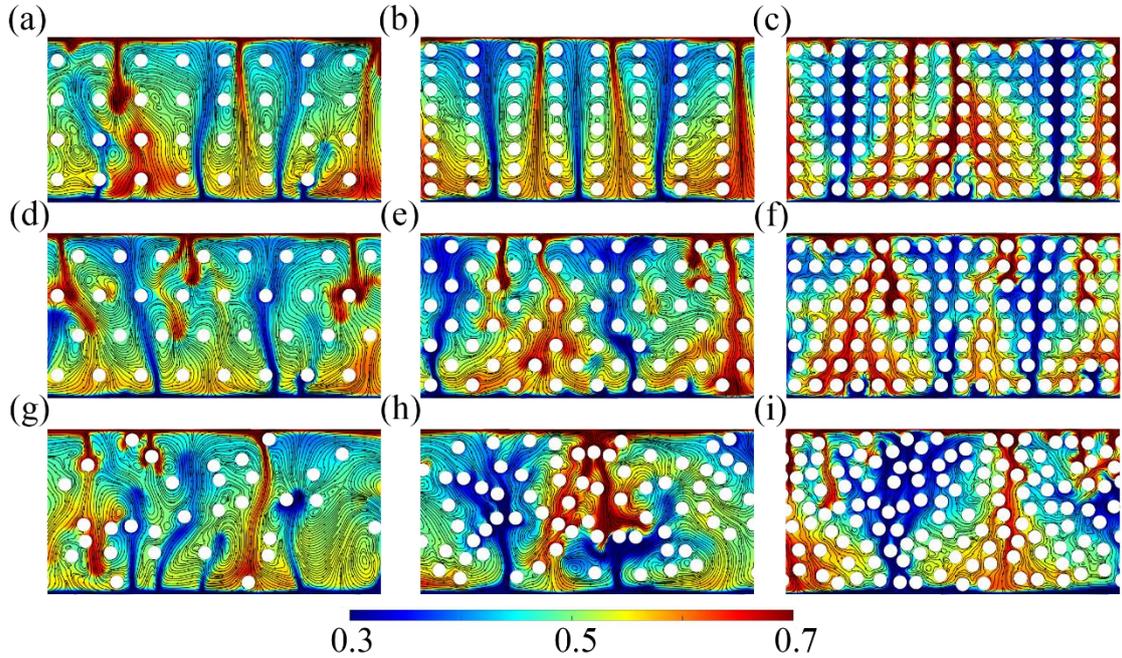

FIG. 13. Typical snapshots of the streamlines superposed on the salinity fields at the quasi-steady stage for $Ra_S = 4 \times 10^7$ with three different arrangements of porous structure (top: lattice arrangement; middle: staggered arrangement; bottom: random arrangement) and three different $\phi$: (a, d, g) $\phi = 0.93$; (b, e, h): $\phi = 0.87$; (c, f, i): $\phi = 0.74$. For a clear presentation of the detailed flow features, only the middle section of the domain is shown. The background color represents the salinity magnitude.

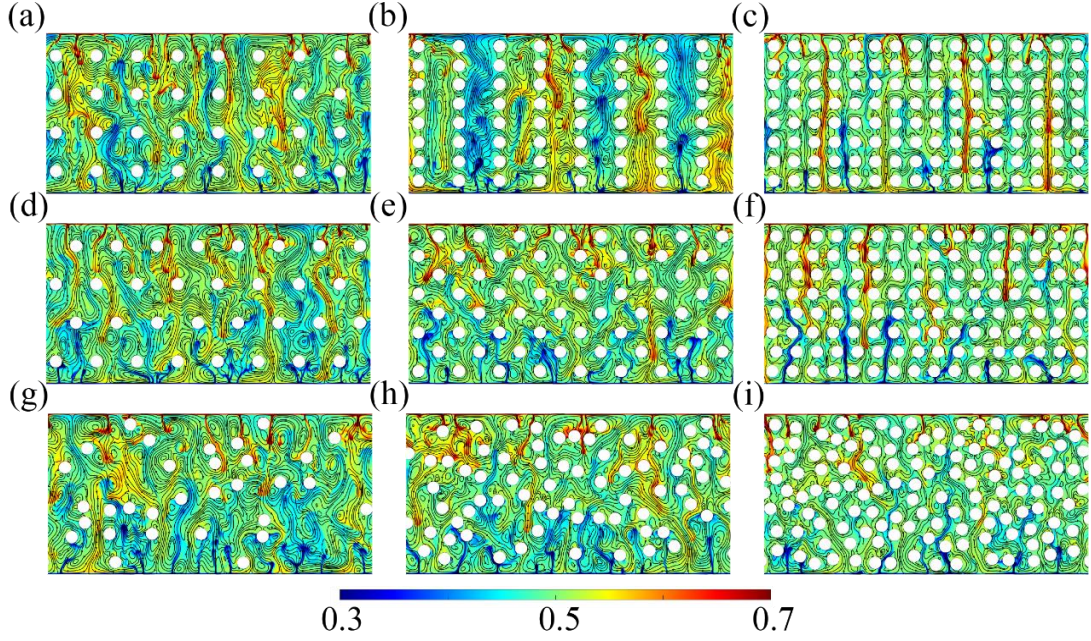

FIG. 14. Typical snapshots of the streamlines superposed on the salinity fields at the quasi-steady stage for $Ra_S = 4 \times 10^9$ with three different arrangements of porous structure (top: lattice arrangement; middle: staggered arrangement; bottom: random arrangement) and three different $\phi$: (a, d, g) $\phi = 0.93$; (b, e, h): $\phi = 0.87$; (c, f, i): $\phi = 0.74$. For a clear presentation of the detailed flow features, only the middle section of the domain is shown. The background color represents the salinity magnitude.

The results of $Nu_S$ and $Re_z$, as shown in Fig. 15, reveal the impact of the porous structure arrangement on flow structures. When the porosity is relatively high ($\phi = 0.93$), the flow patterns are similar among different arrangements. However, as the porosity decreases to 0.87, the impact of the porous structure arrangement becomes more pronounced. While the flow strength remains largely unchanged, the salinity transport in the lattice arrangement is stronger than in the other two arrangements, especially at $Ra_S = 4 \times 10^9$. This confirms that the vertical convection channels play a crucial role in salinity transport for moderate porosities.

At even lower porosities ($\phi$=0.74), the influence of the porous structure arrangement on and $Re_z$ becomes more pronounced as $Ra_S$ increases. For low $Ra_S$, the vertical channels are already destroyed by the porous media; therefore, the transport properties are not sensitive to the arrangement. However, as $Ra_S$ increases, the horizontal scales of the fingered structures decrease, and the vertical channels can form if the porous structure is arranged properly. The more ordered the porous structure, the more vertical channels are allowed to form. As a result, as shown in Figs. 15(c) and 15(f), the salinity transport and flow strength at $Ra_S = 4 \times 10^9$ have the largest values in the lattice arrangement, followed by the staggered arrangement, and the smallest values in the random arrangement. These results highlight how the characteristic scales of the dominant flow structures and their interaction with the porous structure determine the transport properties in the system.

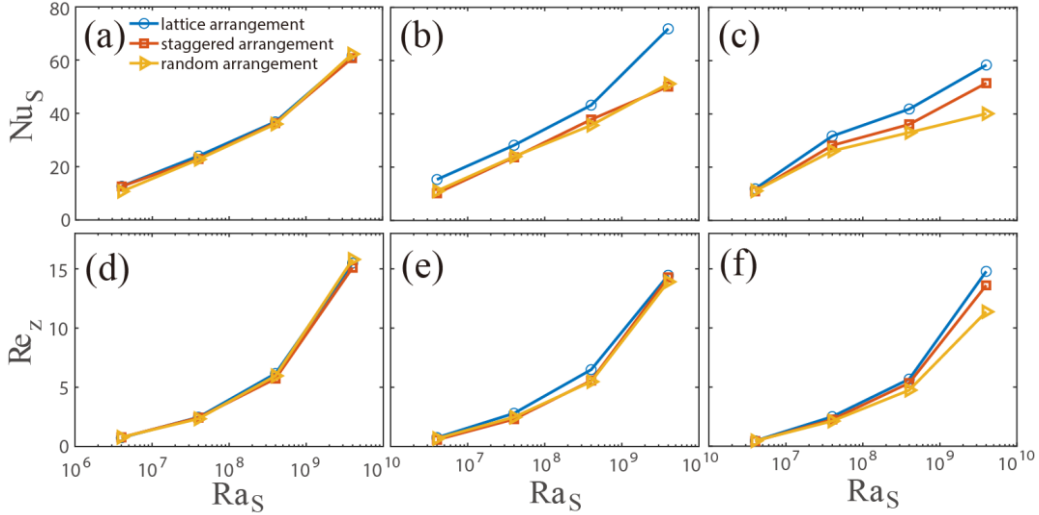

FIG. 15. Comparison of $Nu_S$ and $Re_z$ for different porous structure arrangements under the same porosity $\phi$: (a, d) $\phi = 0.93$, (b, e) $\phi = 0.87$, and (c, f) $\phi = 0.74$.

## IV. CONCLUSIONS

We have conducted a comprehensive study on the behavior of salt finger convection in porous media in the non-Darcy regime. Our study utilized a numerical model based on the lattice Boltzmann method and the immersed boundary method, allowing us to simulate the flow dynamics in the pore scale and its interaction with the porous structure for the first time. Through a series of simulations with a range of porosities ($0.7 \leq \phi \leq 1$) and solute Rayleigh ($4 \times 10^6 \leq Ra_S \leq 4 \times 10^9$), we investigated in detail how the porous media affects the behaviors of salt finger convection. The major findings are as follows:

The development of salt finger convection is delayed by the presence of porous media, and the delay becomes more pronounced as the porosity decreases. The dependencies of solute transport and flow strength on porosity are not straightforward. Specifically, solute transport first enhances with decreasing porosity, and then decreases when the porosity is below a $Ra_S$-dependent value. On the other hand, while flow strength decreases significantly with decreasing porosity at low $Ra_S$, it varies weakly with porosity at high $Ra_S$ and even increases unexpectedly for certain porosities at moderate $Ra_S$.

These complex dependencies are explained by the changes in flow structures. Porous media hinders the movement of fluid and can even destroy dominant flow structures when the pore size is smaller than the width of salt fingers. However, it also weakens mixing and entrainment in the flow, leading to more ordered and coherent fingered structures and the formation of upwelling/downwelling channels that aid in vertical motion and solute transport. The increase in flow coherence can offset the blockage effect on the flow and even enhance transport properties. The interplay between these two competing effects of porous media determines the variation of salinity transport and flow strength with porosity.

Our findings are also applicable to different porous structure arrangements. We found that the impact of the porous structure arrangement is generally stronger for smaller porosities and higher $Ra_S$, owing to the stronger interaction between the fingered structures and porous media.

These findings not only deepen our understanding of convective transport phenomena in porous media, but also have practical applications in fields such as metal processing, agricultural water distribution, and oil recovery techniques.[19–20,57–58]

In this study, we focused on the non-Darcy regime of porous salt finger convection. However, the data shown in Figs. 5(a) and 6(a) indicate a rapid decline when the solute Rayleigh number ($Ra_S$) is less than $4 \times 10^7$ for certain low porosities, which is a hallmark of the transition to the Darcy regime. To accurately determine the transition points and to gain insight into how this transition occurs, further exploration of a wider range of $Ra_S$ values and porosities is necessary. Additionally, the current setup assumes saturation with porous media. It is common for salt finger convection to occur in the scenario where a fluid layer is on top of a porous media layer, such as at the sediment-water interface. Two ongoing studies are addressing these topics and will be reported in the future.

## Statements of Competing Interest

The authors declared that there is no conflict of interest.

## Acknowledgements

We thank Yantao Yang for helpful discussions on this study. This work was supported by the National Natural Science Foundation of China (Grant Nos. 11988102, 11961160719 and 51879086) and the Department of Science and Technology of Guangdong Province (Grant No. 2019B21203001). The authors also would like to thank the computational support from the Center for Computational Science and Engineering at Southern University of Science and Technology.

## Data Availability

The data that support the findings of this study are available from the corresponding author upon reasonable request.